\documentclass[12pt]{iopart}

\usepackage{setstack}
\usepackage{framed}
\usepackage{lscape}
\usepackage{epsfig}
\usepackage{rotating}
\usepackage{cite}
\begin{document}

\title[Exclusion process for particles of arbitrary extension]{
Exclusion process for particles of arbitrary extension: 
Hydrodynamic limit and algebraic properties}

\author{G. Sch{\"o}nherr\footnote[1]{To
whom correspondence should be addressed (g.schoenherr@fz-juelich.de)}
 and G.M. Sch{\"u}tz}

\address{Institut f{\"u}r Festk{\"o}rperforschung, Forschungszentrum 
J{\"u}lich, 52425 J{\"u}lich, Germany}

\begin{abstract}
The behaviour of extended particles with exclusion interaction on a
one-dimensional lattice is investigated. The basic model is called 
$\ell$-ASEP as a generalization of the asymmetric exclusion process 
(ASEP) to particles of arbitrary length $\ell$. Stationary and dynamical 
properties of the $\ell$-ASEP with periodic boundary conditions are derived 
in the hydrodynamic limit from microscopic properties of the underlying 
stochastic many-body system. In particular, the hydrodynamic equation 
for the local density evolution and the time-dependent diffusion constant 
of a tracer particle are calculated. As a fundamental algebraic property 
of the symmetric exclusion process (SEP) the $SU(2)$-symmetry is generalized 
to the case of extended particles.
\end{abstract}


\maketitle

\section{Introduction}
\label{Intro}
There is renewed interest in the investigation of extended particles with 
exclusion interaction. The basic model, which will be referred to as the 
$\ell$-ASEP in the following, is a generalization of the well-studied 
asymmetric simple exclusion process (ASEP) \cite{lig,sch2}. It describes the 
motion of hard rods in one-dimensional discrete space by extended particles 
which move along a lattice according to stochastic hopping dynamics.

The original concept of the $\ell$-ASEP was introduced in 1968 in a paper by 
MacDonald and Gibbs treating protein synthesis \cite{mac1,mac2}. During this 
process, ribosomes move from codon to codon along a m-RNA template, reading 
off genetic information and thereby generating the protein step by step. 
The ribosomes are modeled
as extended particles, which hop stochastically along a chain 
without overlapping each other. Each particle covers several adjacent 
lattice sites to account for the blocking of several codons by a single ribosome.
The attachment of the ribosomes to the m-RNA for the initiation of the protein 
synthesis and their detachment at the point of termination are modeled by 
open boundaries, where particles may enter and exit the lattice with 
rates that differ from the bulk hopping rates.
Using mean-field theory, the authors studied the steady state of this process. 
More recently the time-dependent conditional probabilities \cite{sas}, the 
dynamical exponent \cite{alc99} and the phase diagram of the open system 
have been determined \cite{lak,shaw1,shaw2,shaw3}.

However, the understanding of symmetries of the model and of its
hydrodynamic limit has remained incomplete so far. In \cite{shaw1} a 
hydrodynamic equation is proposed phenomenologically, employing fitting 
parameters, which are matched to simulation data. In the present paper, 
several basic physical and mathematical properties of the 
$\ell$-ASEP with periodic boundary conditions are derived from its microscopic 
dynamics, generalizing a mapping to the zero range process \cite{evan}
and employing quantum Hamiltonian techniques \cite{sch2}. In particular, we 
obtain the hydrodynamic limit governing the 
density evolution of the $\ell$-ASEP on the Euler scale.

The outline of this work is as follows: After introducing the two fundamental 
models ($\ell$-ASEP and zero range process) and reviewing some facts about 
their stationary properties in section \ref{ASEPZRP}, the dynamics of the 
$\ell$-ASEP are studied by two different approaches: In section \ref{tracer}, 
the investigation of the motion of a tagged particle in the framework of the 
quantum Hamiltonian formalism leads to an expression for the average velocity 
and the time-dependent diffusion constant of the tracer particle. 
The velocity term is then confirmed by the general form of a hydrodynamic
equation of the $\ell$-ASEP which is derived in section \ref{HD}. Section 
\ref{SU2} finally exposes the hidden $SU(2)$-symmetry as a fundamental 
algebraic property of the $\ell$-SEP, which arises from the $\ell$-ASEP by 
requiring left/right-symmetric hopping rates.

\begin{figure}
\begin{center}
\epsfig{file=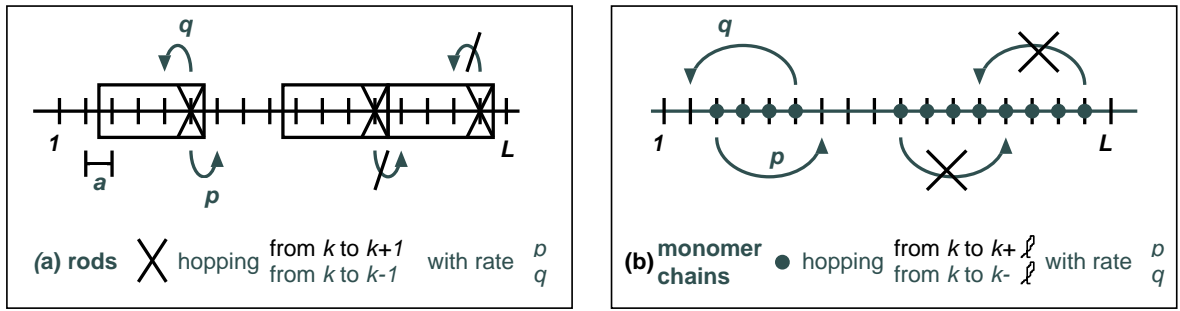, height=\linewidth, angle=270}
\caption{\label{rodspoints}Two approaches to model the $\ell$-ASEP
(left: extended particles which cover $\ell$ lattice sites each and move via 
next-neighbour hopping. right: point particles with certain initial states and 
modified hopping rules.)}
\end{center}
\label{rodspoints}
\end{figure}

\section{$\ell$-ASEP and ZRP}
\label{ASEPZRP}
\hyphenation{For-ma-lism}

\subsection{The $\ell$-ASEP}
\label{ASEP}
The $\ell$-ASEP is the discrete nonequilibrium analogue of a one-dimensional
Tonks gas, including the ASEP as a special case for $\ell=1$. 
$N$ particles are placed on a one-dimensional lattice $S$ consisting of $L$ 
sites $k=1\ldots L$ (Fig. 1(a)). Each particle covers $\ell$ 
adjacent sites. The parameter $\ell$ is an integer number which determines 
the extension of the particles in units of the lattice spacing $a$. In the 
following, $\ell$ will be called the length of a particle. As time proceeds, 
the particles change their locations on the lattice by next-neighbour 
random hopping 
under exclusion interaction. Provided that their right and left neighbour 
sites respectively are not occupied, they move one site to the right with
rate $p$, or one site to the left with rate $q$. The location of 
a particle on the chain is denoted by the location of its right end.

In figure \ref{rodspoints} this model is compared to an equivalent
description (1b), 
where each extended particle is composed of $\ell$ monomers. In the latter 
case, the initial states are restricted to those where particles are grouped 
in $\ell$-tuples at adjacent sites. Hopping processes take place between sites 
which are $\ell$ lattice spacings afar and which enclose $\ell-1$ occupied 
sites in between. This monomer description is useful for the 
application of the quantum Hamiltonian formalism to the $\ell$-ASEP
(see below).

\subsection{Mapping between the $\ell$-ASEP and a ZRP}
\label{ZRP}
One important steady-state property of the $\ell$-ASEP follows directly
from its definition. As sites are always correlated, the $\ell$-ASEP does 
\emph{not} possess a stationary product measure. However, the existence of a 
stationary product measure is an important ingredient in the conventional
derivation of hydrodynamic properties. In order to recover it, the 
$\ell$-ASEP can be mapped onto a different lattice gas model: the zero range
process (ZRP) \cite{spit}.
The ZRP does have a stationary product measure, a fact which is used in 
section \ref{HD} to derive the hydrodynamic equation for the density evolution 
of the $\ell$-ASEP. Furthermore the ZRP-picture will be of help when 
considering the motion of a tagged particle in section \ref{tracer}.
The zero range process is named after the fact that its particles have zero 
interaction range, i.e., there is no exclusion and jump rates do not depend
on the occupation number of the target site. 

In the following, a ZRP will be considered where particle hopping from a site
occupied by $n$ particles occurs with fixed biased rates 
$q$ and $p$ to the left or right respectively.
Like in the case $\ell=1$ \cite{evan}, the $\ell$-ASEP can be mapped onto 
this ZRP by replacing particles by ZRP sites and holes by ZRP particles 
(Fig. \ref{mapping}). 
The appropriate coordinate transformation (compare figure \ref{trafo}) 
between the ZRP with $M$ particles on a lattice of $N+1$ sites 
$j=0,1,\ldots,N$, and the $\ell$-ASEP, having $N$ particles on a lattice of 
$L=\ell N+M$ sites $k=1, \ldots ,L$ is explicitly given by:
\begin{figure}
\begin{center}
\epsfig{file=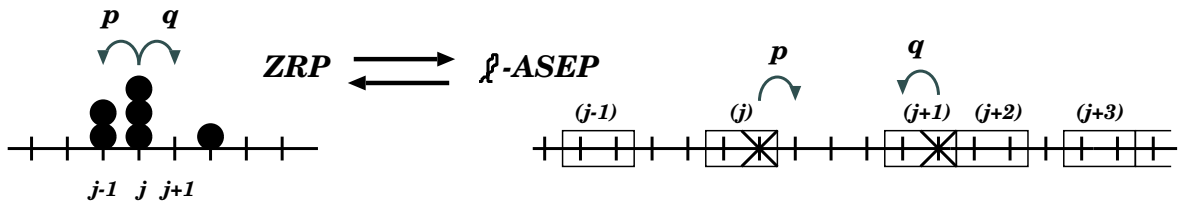, height=\linewidth, angle=270}
\caption{\label{mapping}Mapping between ZRP and $\ell$-ASEP: Lattice sites 
are turned into particles, particles are replaced by holes; the stochastic 
hopping rates $p$ and $q$ are interchanged.}
\end{center}
\end{figure}
\numparts
\begin{equation}
\tilde k = \left( \sum_{j=0}^{\tilde j -1} c_j(t) \right)+\tilde j \ell
\label{trafo}
\end{equation}
where $\tilde k$ is the $\ell$-ASEP lattice site corresponding to a certain 
ZRP site $\tilde j$ and $c_j(t)$ denotes the ZRP particle density at site 
$j$ at time $t$.
The ZRP site $j=0$ is not turned into a particle but determines the position 
of the first $\ell$-ASEP particle. This procedure guarantees the uniqueness 
of the transformation. In the continuum limit, where the lattice constant 
$a$ approaches zero, the discrete coordinates $j$ and $k$ may be replaced 
by continuous variables $y$ and $x$:
\begin{equation}
x=\int_0^{y}\rmd u \, c(u,t) + ly + \frac{a}{2}\left[c(0,t)-c(y,t)\right]+
O(a^2) \, . 
\label{trafox}
\end{equation} 
\endnumparts
The ZRP-density $c$ is related to the $\ell$-ASEP particle density $\rho$ by:
\numparts
\begin{equation}
\rho_{\tilde k}= \frac{1}{c_{\tilde j}+\ell}
\label{dtrafo}
\end{equation}
and
\begin{equation}
\rho(x)=\frac{1}{c(y)+\ell} \, .
\label{traforhox}
\end{equation}
\endnumparts
Whenever speaking of the density $\rho$ in the following, indeed the 
\emph{particle} density, i.e.\ the fraction of particles per lattice unit, 
is referred to, as opposed to the coverage density $\rho^c$ and
hole density $\rho^h$ respectively.
\begin{figure}
\begin{center}
\epsfig{file=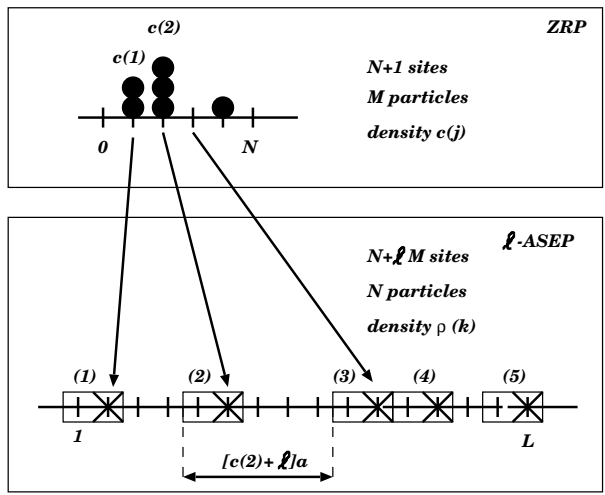, height=8cm, angle=270}
\caption{\label{trafo} Transformation of coordinates and densities: 
ZRP and $\ell$-ASEP quantities are related by a transformation between 
time-independent and time-dependent coordinates; the ZRP-site `zero' provides 
a left boundary of the $\ell$-ASEP lattice (shown here for $N=5,M=6,\ell=2$).}
\end{center}
\end{figure}
\subsection{Stationary state}
\label{stationary}
The stationary state of a zero range process is known \cite{spit}. For a 
periodic system 
all ZRP configurations of the present model are 
equally probable. Due to the existence of the one-to-one mapping between ZRP 
and $\ell$-ASEP configurations on (periodic) lattices of fixed length and 
particle number $(N,L-\ell N)$ and $(L,N)$ respectively, the stationary 
weights of the $\ell$-ASEP must also be distributed equally among all 
configurations on a ring.

The stationary properties of such a system can be deduced from a partition 
function of the form
\begin{equation}
Z=\sum_{N=0}^{N_{max}} z^N Z_N 
\label{Zpartition}
\end{equation} 
where $z$ denotes the fugacity and $Z_N$ is the $N$-particle partition sum. 
$N_{max}$ indicates the maximum number of particles fitting completely on 
$L$ sites. As all states contribute equally, $Z_N$ is given by the number 
of possible different $N$-particle configurations on a lattice of length 
$L$ \cite{lak,busch}:
\begin{equation}
Z_N = \left( \begin{array}{c} L-(\ell -1)N \\ N \end{array} \right)\, .
\end{equation}
The expectation value and the fluctuations of the particle number $N$ are 
calculated from the first and second derivatives of $Z$ with respect to the 
fugacity $z$ using the following standard relations:
\numparts
\begin{eqnarray}
\langle N \rangle=L\rho = \frac{\sum_{N=0}^{N_{max}} Nz^N Z_N}{Z} 
= z\frac{d}{dz}lnZ \\
\langle N^2 \rangle - \langle N \rangle^2      = 
z\frac{d}{dz} \left( z\frac{d}{dz}lnZ \right) \, . 
\end{eqnarray}
\endnumparts
In the hydrodynamic limit the partition sum is approximated by its maximum 
term and a stationary density-fugacity relation may be derived as:
\begin{equation}
z=\frac{\rho (1-(\ell -1)\rho )^{\ell -1}}{(1-\ell \rho )^{\ell}} \, . 
\label{densfug}
\end{equation}
The stationary density fluctuations are given by:
\begin{equation}
\frac{\langle N^2 \rangle-\langle N \rangle^2}{L}=z\frac{d\rho }{dz}
=\rho (1-(\ell -1)\rho )(1-\ell \rho ) \, .
\label{compress}
\end{equation}
For $\ell =1$, equation (\ref{compress}) reduces to the well-known expression 
for the compressibility of the ASEP
\begin{equation}
\kappa_{ASEP}=\frac{\langle N^2 \rangle_{ASEP}-\langle N \rangle_{ASEP}^2}{L}
=\rho (1-\rho )\,.
\end{equation}
The extra factor in (\ref{compress}) may be written as 
$\frac{1}{L}(L-(\ell -1)N)$. This term, ranging between $1/\ell$ and $1$, 
specifies which fraction of the system is formed by holes and particle ends. 
It accounts for the fact that the extended particles, constructed as a chain 
of $\ell$ monomers, are stiff, and that they move simultaneously. The 
fluctuations in the particle number taken per volume of holes and ends 
($L-(\ell-1)N$) are the same for any monodisperse system $S^{(\alpha)}$ 
which contains particles of an arbitrary length $\ell^{(\alpha)}$ provided 
that the particle density and the hole density are fixed a priori.

\section{Motion of a tagged $\ell$-ASEP particle}
\label{tracer}
In experiments, a common procedure to investigate the dynamics of a
diffusive system is to mark a particle and to track its motion. 
In the following, the motion of such a tagged particle is examined with 
analytical tools. The simplest conceivable case is the one of a system 
without interparticle interactions and without the influence of an external 
field. A free particle moving according to symmetric hopping rates on a 
lattice performs a random walk. Starting at time $t=0$ at an initial position 
$x_0$, its location at a time $t>0$ fluctuates stochastically around the 
expectation value $\langle x\rangle \equiv x_0$. An external field, i.e., 
biased hopping rates, causes a particle to move forward into the direction of 
the drive. The expectation $\langle x-x_0\rangle$ of the distance covered by 
this particle is different from zero and proportional to its average velocity. 
Diffusive fluctuations arise again due to the stochastic motion.

If one traces a certain particle of the $\ell$-ASEP, a similar type of motion 
is expected which may be decomposed into a drift term and a diffusion term. 
However, the calculation of the corresponding average velocity $v$ and the 
diffusion constant $D$ is not trivial anymore because collisions cause
time-correlations in the random displacement of the tagged particle.
In order to obtain the average velocity of the marked particle, one counts the 
hopping events that this particle accomplishes starting from a time $t=0$. 
At a time $t>0$, the distance (in units of the lattice spacing) the particle 
has covered is given by the difference of the number of left and right jumps 
it has performed. The average distance is proportional to the average velocity 
$v$ of the tagged particle over that time.

In the ZRP-picture, the average velocity of $\ell$-ASEP particle number $i$ 
corresponds to the difference of leftward and rightward ZRP currents 
$j(i\rightarrow i-1)-j(i-1\rightarrow i)$.
The calculation of these currents is carried out 
conveniently in the quantum Hamiltonian 
Formalism where the master equation of the ZRP is rewritten in terms of a 
Schr{\"o}dinger equation in imaginary time \cite{sch2}:
\begin{equation}
\frac{d}{dt}|P(t)\rangle = -H^{ZRP} |P(t)\rangle \,.
\label{schroedinger}
\end{equation}
The ZRP Hamiltonian $H^{ZRP}$ acts on the state space ${\cal H}$ of the 
probability vectors $|P(t)\rangle$, which are linear combinations of the 
probabilities $P_{\eta}(t)$ to find the ZRP lattice in a state $\eta$ at a 
time $t$ 
\begin{equation}
|P(t)\rangle = \sum_{\eta} P_{\eta}(t)|\eta \rangle \,.
\label{probabilityvector}
\end{equation}
The ZRP Hamiltonian takes a simple form in terms of single-site operators. 
Products of those operators represent transitions between different system 
configurations, i.e. hopping events between adjacent sites. Let $a_j^-$ and 
$a_j^+$ be operators which create and annihilate a particle at site $j$.
The off-diagonal part of the Hamiltonian is formed by the negative sum of all 
such hopping operators, multiplied with the corresponding hopping rates $q$ 
and $p$:
\begin{equation}
\eqalign{H^{ZRP,off}= \sum_{j=0}^{N-1} h^{off}_{j,j+1} \\
h^{off}_{j,j+1}= -qa_j^+a_{j+1}^- -pa_j^-a_{j+1}^+ \, .}
\end{equation}
The diagonal part is deduced from the condition that probability must be 
conserved, i.e. $\langle s|H^{ZRP} = 0$ , $\langle s|$ being the 
$(N+1)$-dimensional constant summation row vector $(1,1,\ldots,1)$. The 
relations 
\begin{equation}
\eqalign{\langle s|a_j^-&= \langle s|\mathbf{1} \\
\langle s|a_j^+&= \langle s|m_j}
\label{createan}
\end{equation}
where $m_j$ is a diagonal operator replacing $a_j^+$ in its action on 
$\langle s|$ yield:
\begin{equation}
\eqalign{H^{ZRP}= \sum_{j=0}^{N-1} h_{j,j+1} \\
h_{j,j+1}= -qa_j^+a_{j+1}^- -pa_j^-a_{j+1}^+ +qm_j +pm_{j+1} \, .}
\end{equation}

\subsection{Extension of the state space}
\label{extension}
Focusing again on the task to keep track of local hopping events, this 
Hamiltonian is modified as follows. In order to create a counting mechanism 
for particles hopping between the chosen sites $i,i-1$, one enlarges the 
tensor state space ${\cal H}$ of the ZRP lattice by an additional infinite 
dimensional subspace 
$\aleph$, whose basis vectors represent
the number $k$ of backward minus forward hops between $i$ and $i-1$. 
Their entries are all zero except the $k^{\rm th}$ counted from the 
``middle'' downwards:
\begin{equation}
\aleph ={\rm Span}\{|k\rangle : k \in {\cal Z}\}
\end{equation}
\begin{equation}
\cdots,|-1\rangle =\left( \begin{array}{c}  \vdots\\ 0\\1\\0\\0\\0\\ 
\vdots\\ \end{array} \right),\quad |0\rangle =\left( \begin{array}{c}  
\vdots\\ 0\\0\\1\\0\\0\\ \vdots\\ \end{array} \right),{} \quad |1\rangle 
=\left( \begin{array}{c}  \vdots\\ 0\\0\\0\\1\\0\\ \vdots\\ \end{array} 
\right),{} \cdots 
\label{alephbasis}
\end{equation}
One modifies the Hamiltonian such that its action on the elements of 
$\aleph$ transfers $|k\rangle $ in $|k \mp 1\rangle$ whenever its action on 
${\cal H}$ leads to a hopping event between the lattice sites $i$ and $i-1$ 
in the ZRP. This is achieved by extra terms
\begin{equation} 
H=H^{ZRP}+ q a_{i-1}^{+}a_{i}^{-}(1-x_{-}) + p a_{i-1}^{-}a_{i}^{+}(1-x_{+}) 
\, .
\label{Hnew}
\end{equation}
All ZRP operators are enlarged to the new state space such that their action 
remains local on ${\cal H}$. $x_-$ and $x_+$ are ladder operators which act 
nontrivially only on $\aleph$, i.e., the zero range operators 
and the counting operators commute. $x_\pm$ are defined by the relations:
\begin{equation}
\eqalign{x_- |k\rangle = |k-1\rangle \\
x_+ |k\rangle = |k+1\rangle \, .}
\label{xplus}
\end{equation}
Finally one constructs an operator $x$ on $\aleph$ for which all 
$|k\rangle$ are eigenstates with eigenvalues k:
\begin{equation}
x|k\rangle =k |k\rangle \, .
\label{x}
\end{equation}
In the chosen basis (\ref{alephbasis}) of $\aleph$, $x$ is represented  as a 
diagonal matrix:
\begin{equation}
x={\rm diag}(\ldots,-2,-1,0,1,2,\ldots)\, .
\end{equation}
The average velocity $v$ and the diffusion coefficient $D$ are derived from 
the long time limits of the expectation value of $x$ and its fluctuations:
\begin{eqnarray}
v = \lim_{t\to\infty} \frac{d}{dt}\langle x \rangle  \label{definitionv}\\
D = \lim_{t\to\infty} \frac{d}{dt}\left[ \langle x^2 \rangle -\langle x 
\rangle^2 \right] \, .
\label{definition}
\end{eqnarray}
The $n^{\rm th}$ moment of the operator $x$ in the stationary state 
$|P^*\rangle $ is given by:
\begin{equation}
\langle x^n \rangle=\langle s|x^n \exp{(-Ht)}|P^*,0\rangle 
\label{moment}
\end{equation}
where
\begin{eqnarray}
|P^*,0\rangle =|P^*\rangle \otimes |0\rangle \nonumber \\
\langle s|=\langle s_1|\otimes \langle s_2|\nonumber \\
\langle s_i|=(...1,1,1...) \in  {\cal H}\backslash \aleph, \;  \aleph \,.  
\nonumber
\end{eqnarray}
While the bra-vector in (\ref{moment}) is the summation vector, adapted to the 
new Hilbert space, the ket-vector is given by the tensor product of 
$|P^*\rangle  \in {\cal H} \backslash \aleph$, characterizing the stationary 
state of the ZRP, and $|0\rangle  \in \aleph$, which sets the counting 
mechanism to zero at the beginning. 

\subsection{General calculation of moments}
\label{pertubation}
In order to calculate the moments  $\langle x^n \rangle $ of the
distribution of the
tagged particle explicitly, the new Hamiltonian (\ref{Hnew}) is formally split into a
non-pertubative term $H_0$ and a pertubative part $V$, where
\begin{eqnarray}
H_0=H^{ZRP}\\
V=q a_{i-1}^{+}a_{i}^{-}(1-x_{-}) + p a_{i-1}^{-}a_{i}^{+}(1-x_{+}) 
\label{V} \, .
\end{eqnarray}
The exponential $\exp{(-Ht)}$ is expanded in a time-ordered Dyson series:
\begin{eqnarray}
\exp{(-(H_0+V)t)}&=\exp{(-H_0t)}[1-\int_0^t \rmd \tau_1 V(\tau_1) \\
 &+\int_0^t \rmd \tau_1 \int_0^{\tau_1} \rmd\tau_2 V(\tau_1) 
V(\tau_2)-\ldots]\nonumber  \\ 
  \nonumber
\end{eqnarray}
where
\begin{equation}
V(\tau)=\exp{(H_0\tau)}V\exp{(-H_0\tau)} \,. \label{Vt}
\end{equation}
For the derivation of the $n^{\rm th}$ moment of $x$, only terms which are at 
most of order $n$ have to be taken into account. All higher orders vanish
identically. Thus
\numparts
\begin{eqnarray}
\fl \langle x \rangle=   \langle s|x \exp{(-Ht)} |P^*,0\rangle  \nonumber \\
 \lo=  \langle s|x \exp{(-H_0t)} |P^*,0\rangle \nonumber \\
   -  \langle s|x \exp{(-H_0t)}\int_0^t\rmd \tau V(\tau) |P^*,0\rangle 
\nonumber \\
  \lo =  \langle x \rangle^{(0)}+\langle x \rangle^{(1)} 
\end{eqnarray}
and
\begin{eqnarray}
\fl \langle x^2 \rangle   =   \langle s|x^2 \exp{(-Ht)} |P^*,0\rangle  
\nonumber \\
\lo=   \langle s|x^2 \exp{(-H_0t)} |P^*,0\rangle \nonumber \\
 -  \langle s|x^2 \exp{(-H_0t)}\int_0^t\rmd \tau V(\tau) |P^*,0\rangle 
\nonumber \\
  +  \langle s|x^2 \exp{(-H_0t)}\int_0^t\rmd \tau_1 V(\tau_1)\int_0^{\tau_1}
\rmd\tau_2 V(\tau_2) |P^*,0\rangle  \nonumber \\
 \lo =  \langle x^2 \rangle^{(0)}+ \langle x^2 \rangle^{(1)}+ \langle x^2 
\rangle^{(2)}\,. \label{expexpansion}
\label{expansion}
\end{eqnarray}
\endnumparts
Setting $\langle x(t=0)\rangle =0$, these expressions reduce to
\numparts
\begin{eqnarray}
\langle x \rangle = \langle x \rangle^{(1)} \label{v}\\
\langle x^2 \rangle  =  \langle x^2 \rangle^{(1)}+ 
\langle x^2 \rangle^{(2)}\,. \label{D} 
\end{eqnarray}
\endnumparts
Explicit expressions for the higher moments in terms of correlation functions
can be obtained analogously, but are not considered below. 
As this strategy of calculating the moments of a random walk driven
by some underlying Markov process does not appear to be widely known,
we stress that
this derivation can be applied to any counting process for any underlying
Markov chain given by some $H_0$. The perturbation $V$ is determined
by the events that increase (multiplied by $x_+-1$) or decrease 
(multiplied by $x_--1$) the counter (particle position) respectively.
In particular, the generating function $\langle \exp{(\alpha x)}\rangle$
can be obtained by replacing $x_\pm \to \exp{(\pm\alpha)}$ in $V$.
This proves that indeed the $m^{th}$ moment is given exactly
by the terms up to same
order in the perturbative expansion. Notice that up to here we did
not use stationarity of the distribution $P^\ast$. The expressions
derived above are valid for any initial distribution.

\subsection[Driven motion: average particle velocity v]{Driven motion: 
Calculation of the average particle velocity $v$}
\label{drift}
First an explicit expression for the average particle velocity 
(\ref{definitionv}) shall be obtained. We present the calculation in some
detail. Equation (\ref{v}) implies:
\begin{eqnarray}
\langle x \rangle &= -  \langle s|x \exp{(-H_0t)}\int_0^t\rmd \tau V(\tau) 
|P^*,0\rangle   \nonumber \\
    &= -  \langle s|x \int_0^t \rmd \tau V(\tau) |P^*,0\rangle  
\end{eqnarray}
where $\exp{(-H_0t)}$ has been absorbed into $\langle s|$. 
Substituting the time-dependent operator $V(\tau )$ by (\ref{Vt}) yields:
\begin{eqnarray} 
\langle x \rangle &=-  \langle s|x \int_0^t \rmd \tau \exp{(H_0t)} 
V \exp{(-H_0t)} |P^*,0\rangle  \nonumber \\
    &= -  \langle s|x \int_0^t \rmd \tau V  |P^*,0\rangle \,.
\end{eqnarray}
The integration is then trivial:
\begin{equation}
  \langle s|x \int_0^t\rmd \tau V  |P^*,0\rangle = t  \langle s|x V  
|P^*,0\rangle \,.
\end{equation}
Inserting the explicit form of $V$ and applying relations (\ref{createan}) 
for the action of the creation and annihilation operators results in the 
simplified expression for the expectation value
\begin{eqnarray}
\langle x \rangle&=& -t [ \langle s_1|q m_{i-1} |P^*\rangle  \langle s_2 
|x (1-x_{-}) |0\rangle \nonumber \\
   & & \qquad +\langle s_1|p m_{i} |P^*\rangle  \langle s_2 |x (1-x_{+}) 
|0\rangle  ] \, .
\label{vorletztes}
\end{eqnarray}
The action of $x_-$ and $x_+$ on $|0\rangle $ and of $x$ on $|-1\rangle ,
|0\rangle , |1\rangle $ is given by relations (\ref{xplus}) and (\ref{x}). 
The expectation of the ZRP-operator $m_k$ in the stationary state is 
calculated from the product measure as
\begin{equation}
\langle s_1|m_k |P^*\rangle  = z 
\end{equation}
where the fugacity $z$ can be identified with the probability for occupancy 
of a site in the stationary state. Equation (\ref{vorletztes}) simplifies to
\begin{eqnarray}
\langle x \rangle &=& -tqz(0+1) -tpz(0-1) \nonumber \\
    &=&t(p-q)z 
\end{eqnarray}
which in terms of the ZRP density $c$ reads
\begin{equation}
\langle x \rangle=t(p-q)\frac{c}{1+c}\,.
\end{equation}
Taking the time derivative and substituting $c$ by the $\ell$-ASEP density 
$\rho$, the average velocity of a tagged particle is obtained as:
\begin{equation}
v=(p-q)\frac{1-\ell \rho}{1-(\ell-1)\rho}\,.
\label{xpunkt} 
\end{equation}

\subsection[Diffusive motion: diffusion constant D]{Diffusive motion: 
Calculation of the diffusion constant $D$}
\label{diffusion}
Equations (\ref{xpunkt}) and (\ref{D}) reduce the derivation of the diffusion 
constant $D$ as defined in (\ref{definition}) to the calculation of the 
temporal derivatives of the first and second order term of the second moment 
of $x$.

The first order term $\langle x^2 \rangle^{(1)}$ is obtained by a 
straightforward calculation similar to the one of $\langle x \rangle^{(1)}$:
\begin{equation}
\frac{d}{dt}\langle x^2 \rangle^{(1)}=-2(p+q)z\,.
\label{firstorder}
\end{equation} 
This is the direct contribution to the diffusion constant that one would
have for Markovian dynamics of the tagged particle, i.e.,
in the absence of any memory effects resulting from the
interaction.

The derivation of the second order 
memory term of $\langle x^2 \rangle $ as given in 
equation (\ref{expexpansion}) is a bit more subtle and requires an 
approximation to yield an explicit expression in terms of the stationary 
density $\rho$. Inserting (\ref{Vt}), absorbing $e^{-H_0t}, e^{-H_0\tau_i}$ 
into $|P^*\rangle $ and into $\langle s|$ and taking the time 
derivative, it simplifies to
\begin{eqnarray}
\frac{d}{dt}\langle x^2 \rangle^{(2)} &=& \frac{d}{dt} \int_0^t\rmd \tau_1 
\int_0^{\tau_1}\rmd \tau_2  \langle s|x^2 Ve^{-H_0(\tau_1-\tau_2)}V 
|P^*, 0\rangle  \nonumber \\
                        &=& \int_0^t\rmd \tau  \langle s|x^2 Ve^{-H_0\tau}V 
|P^*, 0\rangle \, . 
\end{eqnarray}
A straightforward calculation, where $V$ is replaced by (\ref{V}) and 
equation (\ref{createan}) and stationarity of the zero range
distribution is used, turns the memory term 
$\langle x^2 \rangle^{(2)}$ into the integral of the current-current 
correlation function. Translational invariance of the 
system then yields the so far exact expression:
\begin{equation}
\fl \frac{d}{dt}\langle x^2 \rangle^{(2)} = 
\int_0^t\rmd \tau_2 \langle s_1|\left[ q^2 m_{i-1} -2pq m_i + p^2 m_{i+1} 
\right] e^{-H_0\tau}m_{i} |P^*\rangle\,. 
\label{resttask} \vspace*{1cm}
\end{equation}

For symmetric hopping rates ($v=\frac{d}{dt} \langle x \rangle=0$) 
we define the diffusion constant by
\begin{equation}
D=\lim_{t\to\infty} \frac{d}{dt}\langle x^2 \rangle \,.
\end{equation}
Applying the operator relation 
\begin{equation}
\partial_t \langle s_1|n_i = \frac{1}{2}\langle s_1|[m_{i-1}+m_{i+1}-2m_i] 
\, ,
\label{oprelationmn}
\end{equation} 
to expression (\ref{resttask}), summing up (\ref{resttask}) and 
(\ref{firstorder}), and without loss of generality setting $i=0$ the 
time derivative of the second moment $\frac{d}{dt}\langle x^2 \rangle$ 
assumes the form
\begin{equation}
 \frac{d}{dt}\langle x^2 \rangle = \langle n_0(t)m_0(0)
\rangle_{P^*}-\langle n_0\rangle_{P^*}\langle m_0 \rangle_{P*}  \, .
\end{equation}
This exact result is now approximated in terms of the stationary density 
$c=\langle n_0\rangle$ by a linearization of the density at time t:
\begin{equation}
n_0(t)=c+\epsilon_0(t) \, .
\end{equation}
The investigation of the time dependence of the fluctuations $\langle 
\epsilon_x(t) \rangle$ in continuous space which is governed by the 
linearized operator 
equation (\ref{oprelationmn}) leads to a solution for the time-dependent 
correlator in form of a Gaussian function for the initial condition $\langle 
\epsilon_i(0)m_i(0) \rangle= z$. The evaluation at $x=0$ yields:
\begin{equation}
\langle \epsilon_0(t)m_0(0)\rangle = z \frac{1+c}{\sqrt{4 \pi t}} =  
\frac{c}{\sqrt{4 \pi t}}
\end{equation}
In a last step, the stationary ZRP density $c$ is replaced by the 
$\ell$-ASEP density $\rho$, and the time-dependent diffusion constant $D$ 
is obtained as:
\begin{equation}
D =  \frac{1-\ell \rho}{\rho} \frac{1}{\sqrt{4 \pi t}}\,. 
\end{equation}
This is consistent with the well-known subdiffusive behaviour 
\cite{pinc,beij,arra} as observed for the symmetric exclusion process 
($\ell=1$). The novelty of our result is the density dependence of the
amplitude.
Treating the more general case of biased rates $p \ne q$ one expects a 
diffusion constant which depends on the initial state chosen, see
\cite{GodrLuck} for a recent similar result coming from a phenomenological 
approach.

\section{Hydrodynamic equation}
\label{HD}
\subsection{Derivation from ZRP properties}
The one-to-one mapping between ZRP and $\ell$-ASEP is now exploited in order 
to derive the hydrodynamic equation of the $\ell$-ASEP.
In the ZRP there is no exclusion interaction between the particles and a 
hopping event occurs with hopping rates $q$ or $p$, whenever a site is 
occupied by at least one particle. The probability to find site $i$ non-empty 
at time $t$ shall be called $z_i(t)$. In the case of next-neighbour hopping 
and periodic boundary conditions, the ZRP density evolution at any site $i$ 
is described by the master equation:
\begin{equation}
\frac{\partial}{\partial t}c_i(t)=qz_{i-1}(t)+pz_{i+1}(t)-(p+q)z_i(t)\,.
\end{equation}
In the hydrodynamic limit, the lattice constant $a$ approaches zero on a 
coarse-grained scale. Substituting the discrete variable $i$ by a continuous 
variable $y$, the master equation may be expanded into a Taylor series in 
powers of $a$:
\begin{equation}
\frac{\partial c(y,t)}{\partial t}=aB \frac{\partial z(y,t)}{\partial y}+a^2S 
\frac{\partial^2 z(y,t)}{{\partial y}^2}+O(a^3)
\label{zeroequation1}
\end{equation}
where
\begin{equation}
B=p-q\, , \qquad
S=\frac{p+q}{2}
\end{equation}
characterize the contributions from the biased and symmetric part of the 
motion.

Assuming local stationarity at sufficiently long times \cite{kip}, the 
fugacity may be substituted by the stationary density, making use of the 
fugacity-density relation $z(y,t)=\frac{c(y,t)}{1+c(y,t)}$ which is calculated 
from the well-established stationary properties of the ZRP.
Inserting this expression into (\ref{zeroequation1}) yields a continuity 
equation for the ZRP density as a function of coordinate $y$ and time $t$.
The corresponding hydrodynamic equation for the $\ell$-ASEP is calculated by 
applying the transformation rules (\ref{trafox}, \ref{traforhox}). The ZRP 
current $J(c,\partial_y c)$ which is a function of the ZRP density and 
density gradient is thereby mapped onto the $\ell$-ASEP current 
$j(\rho, \partial_x \rho)$, being a function of the $\ell$-ASEP density and 
density gradient with respect to a time-dependent coordinate $x(t)$. 
Carrying out all substitutions in (\ref{zeroequation1}) results in the 
sought-after hydrodynamic equation for the $\ell$-ASEP:

\begin{equation} 
\fl \frac{\partial \rho(x,t)}{\partial t}=-aB\frac{\partial}{\partial x} 
\left[ \frac{\rho(x,t)(1-\ell \rho(x,t))}{1-(\ell -1)\rho(x,t)} \right] 
+a^2S \frac{\partial^2}{\partial x^2} \left[ \frac{\rho(x,t)}{1-(\ell -1)
\rho(x,t)} \right] \, .
\label{lequation}
\end{equation}
Equation (\ref{lequation}) can be rewritten, introducing the `effective 
density'
\begin{equation}
\chi = \frac{\rho}{1-(\ell-1)\rho}
\end{equation}
and the hole density
\begin{equation}
\rho^h= 1-\ell \rho
\end{equation}
as
\begin{eqnarray}
\frac{\partial \rho(x,t)}{\partial t}&=& -aB\frac{\partial}{\partial x}
[\chi \rho^h]+a^2S \frac{\partial^2}{\partial x^2}\chi
\label{lequation2} \\
                                     &=& -\partial_x j(\chi , \rho^h)\,. 
\nonumber
\end{eqnarray}
The form (\ref{lequation2}) of (\ref{lequation}) reminds of the hydrodynamic 
equation for the ASEP in the case $\ell=1$:
\begin{equation}
\frac{\partial \rho(x,t)}{\partial t} = -aB\frac{\partial}{\partial x}
\left[ \rho (1-\rho) \right] +a^2S \frac{\partial^2}{\partial x^2}\rho \, .
\end{equation}
The concept of the effective density was first introduced in \cite{fer} where 
the average velocity of $N$ extended particles of length $\ell$ on $L$ 
lattice sites is proposed to equal $v=1-\chi$.
In terms of $v$, as derived in the previous section, equation 
(\ref{lequation2}) reads:
\begin{eqnarray}
\partial_t \rho &=& - aB\partial_x[v\rho] - a^2S\partial_{xx}v
\label{vZRP} \\
                &=& -\partial_x j(\rho , v)\,. \nonumber
\end{eqnarray}
Equations (\ref{lequation2}) and (\ref{vZRP}) match the result (\ref{xpunkt}) 
for the average particle velocity $v$ as obtained for a tagged particle in the 
previous section.

\subsection{Mapping between ASEP and  $\ell$-ASEP}

A one-to-one mapping between $\ell$-ASEP and its special case of $\ell=1$ can 
be stated explicitly \cite{alc99}.
Let $\rho^1(x',t)$ denote the ASEP particle density as a function of time and 
coordinate $x'$. 
The corresponding transformations are
\numparts
\begin{eqnarray}
x'=x-(\ell -1)\int_0^x \rmd\tilde x\rho(\tilde x,t)+O(a)  \\
\rho^1(x',t)=\frac{\rho(x,t)}{1-(\ell -1)\rho(x,t)}=\chi\,.
\label{atrafo}
\end{eqnarray}
\endnumparts
They prescribe a mapping between states on a lattice of $L'$ sites containing 
$N$ particles of length $1$ and a lattice of $L=L'+(\ell-1)N$ sites containing 
the same number $N$ of particles which have length $\ell$. States which are 
transformed into each other will be called ZRP-equivalent states in the 
following.
Scanning the lattice from left to right and denoting each hole and each 
particle encountered in an ordered sequence ${\gamma_i}$ with ${\gamma_i} 
\in \{\emptyset ,A\}$, where $\emptyset $ represents a hole and $A$ 
represents a particle, yields identical sequences, whenever two states are 
ZRP-equivalent.

\section{$SU(2)$-symmetry of the $\ell$-SEP}
\label{SU2}
For $\ell=1$ and in the case of symmetric hopping rates $p = q =1$ the 
$\ell$-ASEP reduces to the symmetric exclusion process (SEP). One of the 
fundamental algebraic properties of the SEP is the fact that its Hamiltonian 
is $SU(2)$-symmetric \cite{sand}. In the following, the $SU(2)$-symmetry and 
its implications are investigated for the generalized symmetric exclusion
process for particles of length $\ell$ ($\ell$-SEP).

\subsection{Quantum Hamiltonian formalism for the SEP}
\label{HSEP}
In the case $\ell=1$ the quantum Hamiltonian in terms of single-site particle 
creation and annihilation operators $s_k^{\mp}$ and number operators $n_k,v_k$  of the SEP is given by:
\begin{equation}
\eqalign{H^{SEP}&=\sum h_{k,k+1}^{SEP}\\
h_{k,k+1}^{SEP}&=-s_k^-s_{k+1}^+ - s_k^+s_{k+1}^- + v_kn_{k+1} + 
n_kv_{k+1}\, .}
\label{HSEP}
\end{equation}
Choosing a tensor representation with the single-site basis 
\begin{equation}
|A\rangle =\left( \begin{array}{c} 0\\1 \end{array} \right), 
\qquad |\emptyset\rangle =\left( \begin{array}{c} 1\\0 \end{array} \right)
\end{equation}
for one-particle states  $|A\rangle$ and empty states $|\emptyset \rangle$, 
representations of all operators $s_k^{\mp}=(\sigma^x \mp \sigma^y)/2$  
and $n_k=(1-\sigma^z)/2$, $v=(1+\sigma^z)/2$ may be constructed of Pauli 
matrices $\sigma^{x,y,z}$.\\
The $SU(2)$-symmetry of the SEP has useful consequences: The overall particle 
creation and annihilation operators $S^{\mp}= \sum_k s_k^{\mp}$ together 
with the operator $S^3=\sum_k (1/2-n_k)$ form a Spin-1/2 representation of 
the $SU(2)$-Algebra.\\
Utilizing the fact that $H^{SEP}$ commutes in particular with $S^{\mp}$, one 
can show that the local density $\rho_k=\langle n_k\rangle$ $=$ $ \langle 
s|n_k |P(t)\rangle $ satisfies a diffusion equation:
\begin{equation}
\frac{d}{dt}\langle n_k\rangle =- \langle n_k H^{SEP}\rangle = 
\langle n_{k-1}\rangle+\langle n_{k+1}\rangle-2\langle n_k\rangle\,.
\label{diffusion}
\end{equation}
This relation reduces the density evolution of the many-particle problem 
to a single-particle problem (solution of the lattice diffusion
equation) and also implies a correspondingly simple hydrodynamic
limit, viz.\ the diffusion equation. More generally the SU(2)-symmetry implies
that any $k$-point many-particle correlation function can be calculated from 
an associated problem with at most $k$ particles.

Having asserted the existence of a general
one-to-one mapping between ASEP and $\ell$-ASEP (\ref{atrafo}), naturally
the question arises, which operator in the $\ell$-SEP state space follows
satisfies the 
same kind of equation (\ref{diffusion}).

\subsection{Quantum Hamiltonian formalism for the ${\ell}$-SEP}
\label{HlSEP}
The generalized form of the Hamiltonian (\ref{HSEP}) for $\ell \ge 1$ is:
\begin{equation}
\eqalign{H^{\ell-SEP}=&\sum h_{k,k+\ell}^{\ell-SEP}\\
h_{k,k+\ell}^{\ell-SEP}=&-s_k^-n_{k+1}n_{k+2}\cdots n_{k+\ell -1}s_{k+\ell}^+\\
&- s_k^+n_{k+1}n_{k+2}\cdots n_{k+\ell -1}s_{k+\ell}^-\\ 
& + v_kn_{k+1}\cdots n_{k+\ell} \\
& + n_k \cdots n_{k+\ell-1}v_{k+\ell}\,.}
\end{equation}
There are few formal but important physical differences in comparison with the SEP-Hamiltonian (\ref{HSEP}). For instance the action of $s_k^{\mp}$ on any configuration of the lattice $S$ is not equivalent to the creation and annihilation of an extended particle, but it creates or annihilates just a piece of it (one monomer). Furthermore $H^{\ell-SEP}$ is not anymore symmetric under the action of a true particle creation or annihilation operator and such an operator is of no relevance for the $SU(2)$-symmetry. A formula how to construct the operators which are taking over the role of $S^{\mp,3}$ for the $\ell$-SEP as a representation of the Lie-Algebra $SU(2)$ is given in the following.

\subsection{Construction of a new `creation' operator $\tilde S^-$}
\label{creation}
The transformation between SEP and $\ell$-SEP relates states to each other which are ZRP-equivalent.\\
\\
\textit{Definition: Let $|\xi_1'\rangle,|\xi_2'\rangle$ be two arbitrary SEP states obeying}
\begin{equation}
|\xi_2'\rangle=S^-|\xi_1'\rangle
\end{equation}
\textit{and let $|\xi_1\rangle,|\xi_2\rangle$ denote the ZRP-equivalent $\ell$-SEP states. $\tilde S^-$ shall be defined by the relation}
\begin{equation}
|\xi_2\rangle=\tilde S^-|\xi_1\rangle \, .
\end{equation}
$\tilde S^{+,3}$ \textit{are defined analogously.}\\
\\
Due to their construction, $\tilde S^{\mp,3}$ form a representation of $SU(2)$ and the Hamiltonian is symmetric under their action:
\begin{equation}
[H^{\ell -SEP},\tilde S^{\pm,3}]=0\,.
\end{equation}
\begin{figure}
\begin{center}
\epsfig{file=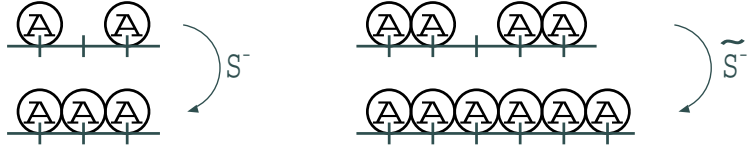, width=10cm, angle=0}
\caption{\label{creation}Creation of particles in a monomer and dimer system, yielding ZRP-equivalent states: $\tilde S^-$ creates dimers on available and  additional volume.}
\end{center}
\end{figure}
While $S^-$ transforms between vectors of a fixed length $2^n$, $\tilde S^-$ maps a vector of length $2^n$ to a vector of length $2^{n+ \ell -1}$. To be able to operate within one state space of fixed dimension, a new model of the SEP and $\ell$-SEP, generalized to two classes of particles $A$ and $B$ will be introduced. For the sake of simplicity of notation the case $\ell=2$ shall be considered first. The results are generalized to $\ell \ge 1$ afterwards.\\
\\
The basis of the new state space $\cal K$ is given by the tensor product states of the single-site basis

\begin{equation}
|0\rangle = \left( \begin{array}{c} 1\\0\\0\\ \end{array} \right), \quad |A\rangle= \left( \begin{array}{c} 0\\1\\0\\ \end{array} \right), \quad |B\rangle = \left( \begin{array}{c} 0\\0\\1\\ \end{array} \right)\,. 
\label{ABbasis}
\end{equation}
The matrix representation of the single-site $A$- and $B$-particle creation, annihilation and number operators in this basis is:

\begin{equation}
\eqalign{s^{A+}=\left( \begin{array}{ccc} 0&1&0\\0&0&0\\0&0&0 \end{array} \right)  \qquad  s^{A-}=\left( \begin{array}{ccc} 0&0&0\\1&0&0\\0&0&0 \end{array} \right)  \qquad n^{A}=\left( \begin{array}{ccc} 0&0&0\\0&1&0\\0&0&0 \end{array} \right)  \\
s^{B+}=\left( \begin{array}{ccc} 0&0&1\\0&0&0\\0&0&0 \end{array} \right) \qquad s^{B-}=\left( \begin{array}{ccc} 0&0&0\\0&0&0\\1&0&0 \end{array} \right)  \qquad n^{B}=\left( \begin{array}{ccc} 0&0&0\\0&0&0\\0&0&1 \end{array} \right) \, .} \vspace*{.5cm} 
\end{equation}
\underline{\bf Monomer space $\cal M$ ($\ell=1$)}\\
\\
The new representation of any monomer configuration $\{\alpha_i \}, \alpha_i \in \{A,0\}$ on the lattice $S'$ of length $L'$ is realized by the following steps:
\begin{itemize}
\item Double the lattice $S'$, adding $L'$ new sites to its right end.
\item Fill the new sites with $B$-particles.
\item Construct the tensor product state representation of the complete chain in the new basis (\ref{ABbasis}).
\end{itemize}
A monomer state which has previously been represented by a $2^{L'}$ dimensional vector $|\xi'\rangle \in \cal H^{\otimes L'}$, is replaced by a $3^{2L'}$ dimensional vector $|\mu \rangle\in \cal K$. All new monomer states $|\mu\rangle $ of arbitrary even dimension form a subspace of $\cal K$: 
\begin{equation}
{\cal M} = \{ |\mu \rangle\} = \{ |\kappa \rangle  \in {\cal K}: a+v=b, \quad \mbox{$B$-particles at right end} \}
\end{equation}
where $a,b$ and $v$ denote the number of $A$-particles, $B$-particles and holes respectively, which are contained in the configuration represented by $|\kappa \rangle $.\\
The dynamics of the monomer system are governed by the monomer Hamiltonian $H^m$. $H^m$ is obtained by substituting all $H^{SEP}$ operators in (\ref{HSEP}) by the ones which are labeled with a superindex $A$ as introduced above. The new vacancy operator $v$ is defined as $v=1-n^A-n^B$. The action of the monomer Hamiltonian $H^m$ is local on the left half of the new monomer system and restricted to $A$-particles and vacancies.\\
\\
\underline{\bf Dimer space ${\cal D}$ ($\ell=2$)}\\
\\
The new representation of a dimer configuration $\{ \beta_i \}, \beta_i \in \{A, \emptyset \}$ on a lattice $S$ of length $L$ is constructed in the same way with the only difference that the number of lattice sites added equals the number $v$ of zeros in $\{ \beta_i \}$. Thus a state previously represented by a $2^L$ dimensional vector $|\xi\rangle$ is replaced by a $3^{L+v}$ dimensional vector $|\delta\rangle  \in \cal K$.
The subspace of $\cal K$, consisting of all such dimer vectors $|\delta\rangle $ is determined by:
\begin{equation}
\fl {\cal D}=\{ \langle \delta \rangle \} = \{ \langle \kappa \rangle \in {\cal K} : v=b,\mbox{ $A$-particles exist pairwise,}
\mbox{ $B$-particles at right end} \}.
\end{equation}
The dimer Hamiltonian $H^d$ for a system of size $L+v$ is given by the $H^{2-SEP}$ for a lattice of length $L$, where all operators except $v$ are labeled with a superindex $A$.\\
\\
\underline{\bf Mapping operators}\\
\\
The new representation enables one to give the mapping between ZRP-equivalent monomer and dimer states explicitly in operator form:
\numparts
\begin{eqnarray}
|\delta\rangle =P^t|\mu\rangle  \\
|\mu\rangle =P|\delta\rangle 
\end{eqnarray}
\endnumparts
where
\begin{equation}
\eqalign{P=\prod_{i=N-1}^1 \left[ \mathbf{1}+(s_N^{B-}s_N^{A+} \prod_{j=N}^{i+1} P_{j,j-1}-\mathbf{1}) n_i^A \right] \\
P^t=\prod_{i=1}^{N-1} \left[ \mathbf{1}+(\prod_{j=i+1}^N P_{j,j-1} s_N^{A-}s_N^{B+}-\mathbf{1})n_i^A \right]\,.}
\end{equation}
The permutation operator $P_{j,j-1}$ permutes two single-site vectors in the tensor product and can be expressed as:
\begin{equation}
\eqalign{\fl P_{ij}=v_iv_j + n_i^An_j^A + n_i^Bn_j^B + s_i^{A-}s_j^{A+} + s_i^{A+}s_j^{A-} + s_i^{B-}s_j^{B+} + s_i^{B+}s_j^{B-} \\ \lo+ s_i^{A-}s_i^{B+}s_j^{B-}s_j^{A+} + s_i^{B-}s_i^{A+}s_j^{A-}s_j^{B+}\,.}
\end{equation}
\begin{figure}
\begin{center}
\epsfig{file=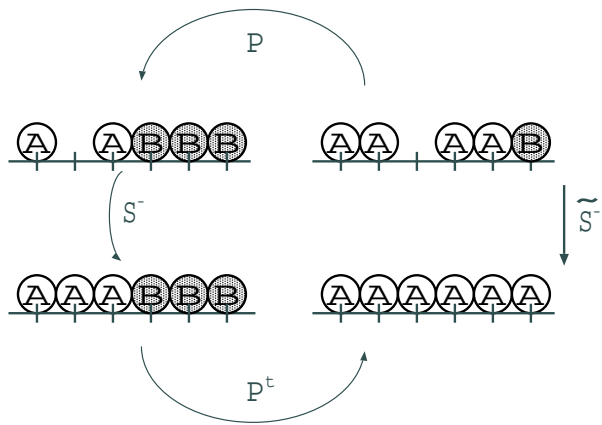, width=8cm, angle=0}
\caption{Action of the mapping operators $P$, $P^t$ on dimer states (right) and monomer states (left), construction of the creation operator $\tilde {S^-}$ from the monomer creation operator $S^-$ in dimer space.}
\end{center}
\end{figure}
The construction of $\tilde S^- \equiv \tilde S^{A^-}$ is now obvious:
Let $|\mu_i\rangle, |\delta_i\rangle,\, (i=1,2)$ be ZRP-equivalent monomer and dimer states with
\begin{equation}
|\mu_2\rangle=S^{A-}|\mu_1\rangle \,.
\end{equation}
Then:
\begin{equation}
\tilde S^{A-} |\delta_1\rangle = |\delta_2\rangle = P^t |\mu_2\rangle= P^t S^-|\mu_1\rangle = P^t S^- P |\delta_1\rangle \,. 
\end{equation}

\subsection{Diffusion equation for $\ell$-SEP operators}
\label{diffusioneq}
All dimer operators $O^{\delta}$ can be constructed as $P^tO^{\mu}P$ from monomer operators $O^{\mu}$. The operator $\tilde n_k = P^t n_k P$ takes over the role of the monomer number operator: $\tilde n_k$ is the quantity which fulfills a diffusion equation with respect to the dimer Hamiltonian $H^d$:
\begin{eqnarray}
\langle \tilde s|\tilde n_k H^d &=&  \langle s|P P^t n_k P H^d \nonumber \\
&=& \langle s|n_k H^m P  \nonumber \\ 
&=&  \langle s|(n_{k-1}-n_{k+1}+2n_k) P \nonumber \\
&=& \langle \tilde s|P^t(n_{k-1}-n_{k+1}+2n_k)P \nonumber \\
&=& \langle \tilde s|(\tilde n_{k-1}- \tilde n_{k+1}+2 \tilde n_k)\,.
\end{eqnarray}
The validity of $PH^d |\delta\rangle =H^mP |\mu\rangle $ in $\cal K$ follows from the construction of the mapping. However, $\tilde n_k$ is not diagonal in the chosen basis of $\cal K$, and it is hard to draw conclusions as to its expectation value. Therefore a diagonal operator $Q_k$ is constructed, which equals $\tilde n_k$ in its action on $\langle  \tilde s| = \langle s|P$. \\
The operator $\tilde n_k$ picks those dimer states from $\langle  \tilde s|$, which are ZRP equivalent to the ones, $n_k$ picks from $ \langle  s|$. The action of an operator $Q_k$ which replaces $\tilde n_k$ thus cannot be local on the site $k$ but must involve several lattice sites. Its action on a certain configuration depends on the number of particles and its label $k$.\\
\\
$Q_k$ can be expressed in terms of diagonal matrices $Q_k^{\alpha}$ only: 
\begin{equation}
Q_k=\sum_{\alpha=0}^{k-1}Q_k^{\alpha}
\end{equation}
where
\begin{equation}
\fl \eqalign{\mbox{for } \alpha=0: Q_k^0 = \prod_{s=1}^{k-1}v_s n_k  \\
\mbox{for } \alpha \ge 1: Q_k^{\alpha} = \sum_{r \in {\cal R}^{\alpha}} \left( \prod_{i=1}^{\alpha} n_{r_i} \right) \left( \prod_{s \in S_r} v_s \right) n_{k+ \alpha}  \label{alpha}\\
{\cal R} ^{\alpha} = \left\{ r = (r_1, \cdots, r_{\alpha}): [r_i \in \{ 1, \cdots, \alpha + k - 2 \}] \wedge  [r_i+2 \le  r_{i+1}] \right\} \\
S_{\mathbf{r}} = \{ 1, \cdots , \alpha +k-1 \} \backslash [ \cup_{i=1}^{\alpha} \{r_i,r_{i+1} \} ] \,. }
\end{equation}
The local monomer number operator $n_k$ picks all states from $ \langle  s|$ where site $k$ is occupied. $Q_k$ instead chooses all such dimer states from $\langle  \tilde s|$ where the $(\alpha+1)^{\rm th}$ dimer, counted from the left, covers sites $k+ \alpha$ and $k+\alpha+1$, and where $\alpha$ is a number between $1$ and $k$.
Thus each $Q_k^{\alpha}$ sums up all possible configurations of placing $\alpha$ dimers ($ \alpha$ $A$-particle-pairs) and $k-1-\alpha$ vacancies on the first $k+\alpha-1$ sites. The position of the left $A$-particle of dimer number $i$ is chosen with the element $r_i$ of the vector $r$. Its elements must appear in certain configurations due to the pairwise arrangement and exclusion interaction of the $A$-particles.\\
\\
It is straightforward to generalize $Q_k$ to the case of particles of arbitrary length $\ell \ge 1$. The case of $\alpha \ge 1$ in (\ref{alpha}) is substituted by the general expression:

\begin{equation}
\fl \eqalign{Q_k^{\alpha} = \sum_{r \in {\cal R}^{\alpha}} \left( \prod_{i=1}^{\alpha} n_{r_i} \right) \left( \prod_{s \in S_r} v_s \right) n_{k+(\ell-1) \alpha} \\
{\cal R} ^{\alpha} = \{ r = (r_1, \cdots, r_{\alpha}): [r_i \in \{ 1, \cdots, k+ (\ell-1) \alpha - \ell  \}] \wedge  [r_i+\ell \le  r_{i+1}] \}  \nonumber \\
S_{\mathbf{r}} = \{1,\cdots,k+(\ell -1)\alpha - \ell\}\backslash [\cup_{i=1}^{\alpha} \{r_i,r_i+1, \cdots r_{i+(\ell-1)}\}]\,. \nonumber}
\end{equation}

\section{Summary and Conclusions}
The purpose of this work was to investigate the properties of extended interacting particles, moving stochastically on a one-dimensional lattice. The main results can be summarized as follows:
One-to-one mappings between $\ell$-ASEP, ASEP and a certain class of ZRP have been stated explicitly. It has turned out very useful to exploit those transformations in order to derive basic properties of the $\ell$-ASEP, in particular the time-dependent diffusion constant for a tracer particle and the hydrodynamic equation for the local density evolution. A tagged $\ell$-ASEP particle shows the type of subdiffusive behaviour as known for the case $\ell=1$ of the ASEP. The extension of the particles as a new feature becomes manifest in the prefactor of the diffusion constant, which has been calculated as a function of the particle density. The methods applied are suitable for a generalization to any process which can be mapped onto some zero range process. As a most important outcome of the mapping, the hydrodynamic equation of the $\ell$-ASEP has been deduced from microscopic properties of the discrete system. The resulting nonlinear and convex current-density relation (for finite $\ell$) is qualitatively similar to that of the ASEP, but it also shows some new features: It is asymmetric for $\ell \ne 1$. The symmetry of particle and hole density is broken. The hydrodynamic equation has a natural form for $\ell \ge 1$ if expressed in terms of the particle density and of a generalized average particle velocity. All results obtained for the $\ell$-ASEP also hold for a polydisperse system of particles of arbitrary length where the length parameter $\ell$ must be replaced by an average length $\bar \ell = \frac{1}{N} \sum_{i=1}^N \ell_i$.\\ 
In the case of $\ell=1$, it is known how to link some hydrodynamic properties of the ASEP to the algebraic structure of the stochastic many-body system. Especially for the case of symmetric hopping rates (SEP), the $SU(2)$-symmetry has proved a valuable attribute. In this work, the $SU(2)$-symmetry has been established for the case of extended particles ($\ell$-SEP). A formalism has been introduced in which all SEP-operators may be generalized to $\ell$-SEP operators under the condition of ZRP-equivalence.\\

\end{document}